\documentclass[useAMS,usenatbib]{mn2e}

\usepackage{txfonts}
\usepackage[dvips]{graphicx}
\usepackage{epsfig}
\usepackage{subfigure}
\usepackage{natbib}
\bibpunct{(}{)}{;}{a}{}{,}

\newcommand{\sref}[1]{Section \ref{#1}}
\newcommand{\fref}[1]{Fig. \ref{#1}}
\newcommand{\tref}[1]{Table \ref{#1}}
\newcommand{\change}{\rm}
\newcommand{\changeend}{\rm}

\title[Future virialized structures]%
{Future virialized structures: An analysis of superstructures in
SDSS-DR7}

\author[Luparello et al.]
  {H.~Luparello$^{1}$, M.~Lares$^{1}$, D.~G.~Lambas$^{1}$ and N.~Padilla$^{2}$\\
 $^{1}$Instituto de Astronom\'{\i}a Te\'{o}rica y Experimental
  (CONICET-UNC). Observatorio Astron\'{o}mico de C\'{o}rdoba, 
  Laprida 854, X5000BGR, C\'{o}rdoba, Argentina\\
 $^{2}$Departamento de Astronom\'{\i}a y Astrof\'{\i}sica, Pontificia
 Universidad Cat\'olica de Chile, Santiago, Chile}

\date{Released 2002 Xxxxx XX}
\pagerange{\pageref{firstpage}--\pageref{lastpage}} \pubyear{2002}

\def\LaTeX{L\kern-.36em\raise.3ex\hbox{a}\kern-.15em
    T\kern-.1667em\lower.7ex\hbox{E}\kern-.125emX}

\begin{document}
\label{firstpage}
\maketitle

\begin{abstract}
\change{}
We construct catalogues of present superstructures that, according to
a $\Lambda CDM$ scenario, will evolve into isolated, virialized structures in
the future.
We use a smoothed luminosity density map derived
from galaxies in SDSS-DR7 data and separate high luminosity density
peaks.
The luminosity density map is obtained from a volume--limited
sample of galaxies in the spectroscopic galaxy catalogue,
within the SDSS-DR7 footprint area and in the redshift range
$0.04<z<0.12$.
Other two samples are constructed for calibration and
testing purposes, up to $z=0.10$ and $z=0.15$.  
The luminosity of each galaxy is spread using an Epanechnikov kernel 
of 8Mpc/h radius, and the map is constructed on a 1 Mpc/h cubic cells grid.
Future virialized structures (FVS) are identified as regions with overdensity above
a given threshold, calibrated using a $\Lambda CDM$ numerical
simulation, and the criteria presented by \citet{duenner_limits_2006}.
We assume a constant mass--to--luminosity ratio and impose the further
condition of a minimum luminosity of $10^{12}L_{\odot}$.
According to our calibrations with a numerical simulation, 
these criteria lead to a negligible contamination by less overdense (non FVS) superstructures.
We present a catalogue of superstructures in the SDSS-DR7 area within redshift $0.04<z<0.12$
 and test the reliability of our method by studying
different subsamples as well as a mock catalogue.
We compute the luminosity and volume distributions of the superstructures
finding that about 10\% of the luminosity (mass) will end up in future
virialized structures.
The fraction of groups and X--ray clusters in these superstructures is higher 
for groups/clusters of higher mass, suggesting that future cluster mergers will
involve the most massive systems.
We also analyse known structures in the present Universe and
compare with our catalogue of FVS.
\changeend{}
\end{abstract}

\begin{keywords}
large scale structure of the universe - statistics - data analysis
\end{keywords}



\section{Introduction} \label{S_intro}

Structures in the universe are the result of a hierarchical process of
accretion dominated, in almost all scales and during almost their
entire history, by gravity.
The first attempt to investigate the evidence for the existence of the
so called second-order clusters of galaxies was made by
\citet{shapley_review_1961}, although there had been previous
suggestions of the presence of large structures.
\citet{shapley_extension_1930} suggested the presence of a large
galaxy system in the Coma-Virgo region.
Also, conglomerates composed by several clusters of galaxies observed
in plates of the Lick Astrographic Survey were reported by
\citet{shane_distribution_1954} and \citet{shane_distribution_1956}.
By analyzing the distribution of rich clusters identified on the
National Geographic Society-Palomar Observatory Sky Survey, Abell
pointed out that the clusters did not appear to be distributed
randomly over the sky, but forming associations of matter on greater
scales.
Different authors adopted a variety of methods to search for groups of
clusters and the term supercluster became widely used.
Abell used counts in cells to statistically test the distribution of
clusters, and endorsed the existence of superclusters of order
\mbox{50 $h^{-1}Mpc$} in size.
The first attempts to study superclusters on a statistical basis were
performed by linking Abell cluster positions
\citep{zucca_allsky_1993,einasto_superclustervoid_1997}.
Later, the accomplishment of wide--area surveys of galaxies with
spectroscopic follow up, such as Las Campanas Redshift Survey
\citep{LCRS}, the 2-degree Field Galaxy Redshift Survey
\citep[2dFGRS,][]{colless_2df_2001} and the Sloan Digital Sky Survey
\citep[SDSS,][]{stoughton_sloan_2002}, allowed for the identification
of superclusters directly from the large-scale galaxy distribution.
\citet{einasto_superclusters_2007} identified superclusters in the
2dFGRS using a density field method,
\change
and recently \citet{costaduarte_morphological_2010}, studied the morphology
of superclusters of galaxies in the SDSS.
The largest catalogue of superclusters has been constructed by \citet{liivamagi_sdssdr7_2010},
who implemented the density field method on the SDSS--DR7 main and LRG samples.
\changeend                            
In all cases, superclusters are operationally defined as objects
within a region of positive galaxy density contrast and thus are
subject to a certain degree of arbitrariness in the parameter
selection.

Within the $\Lambda CDM$ Concordance Cosmological Model, an accelerated
expansion dominates the present and future dynamics of the universe
and thus determines the nature of gravitationally bound structures.
Therefore, an alternative definition of these large-scale structures
may be derived from the properties of
overdense regions in the
present-day universe, that will become bound and virialized structures
in the future.
Thus, under the assumption that luminosity is a somewhat unbiased
tracer of mass on large scales, the integrated luminosity density of
galaxies is commonly used as an indicator of mass density.

The cosmological evolution of the large scale structure has
implications on the spatial distribution, frequency and properties of
superclusters and the galaxies they contain.
Due to this relationship, superclusters can be used as cosmological
probes, and their study is oriented to constrain models and describe
the formation of superstructures on a cosmological scale.
Supercluster properties have been used to discriminate between
cosmological models, favoring the standard cosmological model in most
cases \citep{basilakos_pscz_2001,peacock_measurement_2001}.
On the other hand, there are claims of structures that are too massive
or formed too early according to the standard model
\citep{baugh_2df_2004}.

\change
First attempts to determine the characteristic scales of spatial inhomogeneities
in the universe were made by \citet{broadhurst_1990} and \citet{einasto_sclvoidnet_1994}.
By studying the distribution of rich clusters of galaxies from the Abell-ACO catalogue,
\citet{einasto_sclvoidnet_1994} confirmed a $110-140h^{-1}Mpc$ scale
in the supercluster--void network.
Given the observed regular pattern of superclusters and voids, \citet{frisch_1995} investigated
the properties of the initial power spectrum giving rise to these large--scale fluctuations. 
The authors found that the supercluster--supervoid network forms in a very early stage of the 
evolution of the Universe from large--scale density fluctuations, and are defined by
the scale of the maximum of the power spectrum.
\changeend

The large scale structure of the universe is often described as a
supercluster-void network \citep{shandarin_morphology_2004}, and
superclusters are closely related to filaments
\citep{gonzalez_automated_2010, murphy_filamentary_2010} and voids
\citep{einasto_structure_1986,einasto_superclustervoid_1997,park_void_2007,
platen_alignment_2008,icke_voids_1984}.
Such filaments may also play an important role in the process of
structure formation.  \citet{porter_star_2008} find that star
formation is significantly enhanced when galaxies fall into clusters
along supercluster filaments.
Superclusters have also been studied as hosts of Ly$\alpha$ absorbers
\citep{stocke_local_1995,penton_local_2002} and are known to produce
signatures in the Cosmic Background Radiation
\citep{granett_dark_2009,
dolag_imprints_2005,genova_observations_2008,flores_sunyaev_2009,genova_study_2010}.

The aim of this work is to present and analyse catalogues of
superstructures that will evolve into virialized systems.
Using the theoretical framework of \citet{duenner_limits_2006} in a
$\Lambda CDM$ scenario and by calibration with numerical simulations
we analyse volume limited samples of galaxies from the SDSS--DR7.

This paper is organized as follows.
In section \sref{S_data} we describe the data used in the computation
of the density field.
The methodological description of the procedures implemented to obtain
the catalogue of future virialized structures (hereafter FVS) is
addressed in \sref{S_method}, along with a brief description of the
previous methodology employed in the search of superclusters.
The catalogue of FVS is presented in \sref{S_catalog}, where we also
discuss some of their properties.
In \sref{S_SCSvsFBS} we compare known superclusters with our
identified structures, and in section \sref{S_SCSinFBS} we study the
frequency of clusters and groups of galaxies in future virialized
structures.
Our conclusions are summarized in \sref{s_conclusus}.
Throughout this paper, we adopt a concordance cosmological model
($\Omega_{\Lambda}=0.75$, $\Omega_{matter}=0.25$) in the calculation
of distances.
%

\section{Data and Samples} \label{S_data}

Since FVS are extended regions  of several tens of Mpc in size, a
large volume of space has to be surveyed in order to sample a
representative population of these objects.
The Sloan Digital Sky Survey \citep[SDSS,][]{stoughton_sloan_2002} is
the largest photometric and spectroscopic survey carried out so far,
covering an area of almost a quarter of the sky with a limiting
magnitude that makes the construction of complete samples of several
hundred Mpcs possible.
Several catalogues have been made public by successive releases since
Early Data Release \citep{stoughton_early_2002}.
The latest release of the spectroscopic catalogue
\citep[DR7,][]{abazajian_seventh_2009} comprises $929\,555$ galaxy
spectra within a footprint area of 9380 sq. deg.
The limiting apparent magnitude for the spectroscopic catalogue in the
r-band is $17.77$ \citep{strauss_spectroscopic_2002}, although we use
a more conservative limit of $17.5$ to ensure completeness.
We also limit the sample to galaxies fainter than $r=14.5$, since
saturation effects in the photometric pipeline does not secure
completeness below that limit.  These limits were adopted taking into
account the analysis of image quality and efficiency detection of the
SDSS
\footnote{http://www.sdss.org/dr7/products/general/target\_quality.html}.

Given the dependence of the luminosity density field on the sample of
galaxies, we define three complete samples, with different cuts in
luminosity (\tref{T_samples}).
The closest sample comprises galaxies brighter than \mbox{M$_r=-20.5$}
up to \mbox{$z=0.1$}.
This is the sample with the faintest luminosity limit, and it will be
mainly used to explore the effects of the luminosity cut in the
detection of superstructures.
This sample will be referenced as S1, and their characteristics are
summarized in \tref{T_samples}.
A larger volume limited sample, referenced as S3, contains all
galaxies brighter than \mbox{M$_r=-21.0$} to $z\le 0.15$.
An intermediate redshift sample in the range $0.04<z<0.12$ comprises
$89513$ galaxies.
This sample, S2, containing galaxies with \mbox{M$_r<-20.47$} will be
analised into detail in Sections \ref{S_catalog}, \ref{S_SCSvsFBS} and
\ref{S_SCSinFBS}.

\begin{table*}
\centering
\begin{tabular}{@{\extracolsep{\fill}}ccccccccc}

\hline 
Sample & $z_{max}$ & $D_{max} [h^{-1}\,Mpc]$  & $M_r^{lim}$ & Volume
$[10^7(h^{-1}\,Mpc)^3]$ & $N_{gal}$ & $\bar{\rho}_{lum} [10^8L_{\odot}/Mpc^3]$ & $F$ & Corrected $\bar{\rho}_{lum} [10^8L_{\odot}/Mpc^3]$\\

\hline 
S1  & 0.10 & 293.92 & -20.05 & 1.85 & 94271 & 0.80 & 2.11 & 1.68   \\
S2  & 0.12 & 351.34 & -20.47 & 3.17 & 89513 & 0.58 & 2.98 & 1.73   \\
S3  & 0.15 & 436.55 & -21.00 & 6.01 & 62344 & 0.29 & 5.66 & 1.64   \\
\\                                                 
S2c & 0.10 & 293.92 & -20.47 & 1.85 & 51188 & 0.56 & 2.98 & 1.73   \\
S3c & 0.10 & 293.92 & -21.00 & 1.85 & 17507 & 0.27 & 5.66 & 1.64   \\
\\                                                 
$M_{Rsp}$&0.12& 351.34 & -20.47 & 3.17 & 106604 & 0.75 & 2.98 & 2.23 \\
$M_{Zsp}$&0.12& 351.34 & -20.47 & 3.17 & 106722 & 0.75 & 2.98 & 2.23 \\

\hline 
\end{tabular}
\caption{%
Galaxy samples in the SDSS--DR7. In all cases: $z \geq 0.04$, 
and the apparent magnitude in the
r--band is in the range $14.5\leq r \leq 17.5$.  The mean luminosity
density $\bar{\rho}_{lum}$ is computed using volume limited samples,
each containing $N_{gal}$ galaxies.
The correction factor $F$ (Eq. \ref{F_correctionL}) and the resulting 
mean luminosity $\bar{\rho}_{lum}$ of each sample are indicated in 
the table.}
\label{T_samples}
\end{table*}

In order to test procedures and results, we use the Millennium
numerical simulation \citep{springel_simulations_2005} of a $\Lambda
CDM$ concordance model, performed on a cubic box of \mbox{500 $h^{-1}$
Mpc} side.
A semi--analytical model of galaxy formation
\citep[GALFORM,][]{bower_flip_2008} collects information from the
merger trees extracted from the simulation, and generates a population
of galaxies within the simulation box.
We use the semi--analytic galaxy catalogue in the full box of the
simulation to test and set the parameters involved in the
identification of FVS.
We constructed a mock catalogue from the semi--analytic galaxies
following the geometry of the SDSS--DR7 footprint area and reproducing
the dilution in the number of galaxies with redshift.
In order to test the effect of peculiar velocities in the
identification of FVS, we defined a sample of galaxies in real--space
$M_{Rsp}$ and a sample of galaxies in redshift space $M_{Zsp}$, also
described in \tref{T_samples}.

We use AB magnitudes and apply k+e corrections in the rest frame at
$z=0.1$, as defined in \citep{blanton_estimating_2003}.
%
From these magnitudes, we calculate the luminosity of each galaxy, in
the r--band, as $L=L_{\odot} \times 10.^{(0.4 \times
(M_{\odot}-M_r))}$.
The mean luminosity density $\bar{\rho}_{lum}$ of each sample is
computed with this luminosity, and is also listed in \tref{T_samples}. 
We will also study the properties of groups and clusters of galaxies
within FVS.
To this end, we use a catalogue of galaxy groups in the SDSS--DR7
compiled by \citet{zapata_influence_2009}.
This catalogue is obtained using an adaptative Friends-of-Friends
method.
Virial masses are computed from the line of sight velocity dispersion
of galaxies in each group ($\sigma_v$) and the virial radius
($R_{vir}$), using the virial theorem
\citep[see][]{zapata_influence_2009}:

\begin{equation}
M_{vir} = \frac{3 \, \sigma_v^2 \, R_{vir}}{G},
\label{E_Mvir}
\end{equation}

\noindent where $R_{vir}$ is estimated as in
\citet{merchan_galaxy_2005}.
The catalogue comprises 83784 groups with at least 4 members, and is
limited to redshift $z<0.12$.

We have also analysed a catalogue of X--ray selected clusters
\citep{popesso_rass_2004} in order to study the presence of large
gravitational potential wells in FVS.  
This sample comprises $114$ clusters with X--ray emission, and is
based on the ROSAT All Sky Survey (RASS) and the Sloan Digital Sky
Survey (SDSS).
The total luminosity in the ROSAT band is available for each cluster,
and is used as a proxy for the cluster mass \citep{rykoff_LXM_2008}.
We use this luminosity in \sref{S_SCSinFBS} to explore the variations
with X--ray luminosity of the fraction of clusters belonging to FVS.

\section{Method} \label{S_method}

\begin{figure}
   \centering 
   \includegraphics[width=0.5\textwidth]{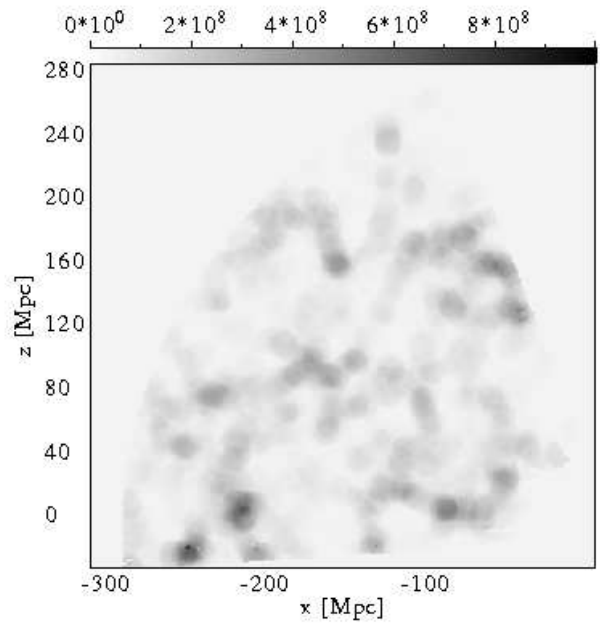}
   \caption{A slice of the luminosity density map for the S2 sample.
   The cartesian system used is associated with the equatorial
   coordinate system, with the z-axis towards the north celestial pole. The
   smoothing kernel has an Epanechnikov shape, with a size of \mbox{8
   $h^{-1}Mpc$}}
   \label{f_Ldensmap} 
\end{figure}
 

 \subsection{Previous analyses} \label{SS_OthersMethod}
 
Several procedures have been proposed to construct a reliable catalog
of superstructures, although they are based on different criteria and
methods.
The main issue related to the comparison of superstructure catalogues
is that the definition of superstructures depends on the specific
choice of a number of free parameters.
\citet{einasto_superclustervoid_1997} use a percolation algorithm to link
clusters of galaxies from the Abell-ACO catalogue, and derive a
catalogue of 220 superclusters of rich clusters within $z=0.12$.
The authors percolate clusters lying within a given radius from each
other, defining these systems of clusters as superclusters.
They find that a percolation radius of \mbox{24 $h^{-1}Mpc$} is
convenient to detect the largest, but still relatively isolated,
systems of clusters.
The main results from the supercluster catalogue are not, according to
the authors, very sensitive to the exact value of the percolation
radius.
\citet{einasto_optical_2001} updated the catalogue with a newer
version of the Abell catalogue, incorporating to the analysis a sample
of X--ray clusters; this also allowed them to compare superclusters
derived from different samples of clusters.
They find that both types of clusters generate superstructures that
represent the large scale structure in a similar way.
Also, they find that X--ray clusters not belonging to superclusters
surround the Southern and Northern Local supervoid, or are located in
filaments between superclusters.
Although the authors find a strong signal indicating that the fraction
of X--ray clusters in superclusters increases with supercluster
richness, there is no correlation between the X--ray luminosity of
clusters and their host supercluster richness, quantified as the
number of member clusters.
\citet{einasto_superclusters_2007} return to the problem of defining
the largest isolated structures in the universe, and use data from the
2dFGRS \citep{colles_2dF_2001} to assemble a catalogue of
superclusters.
The authors find that the most effective method to perform a
supercluster search is the density field method which consists
 in obtaining a smoothed luminosity density field in redshift--space
 from the galaxy catalogue.
%
The luminosity, on large scales, is supposed to follow the
distribution of matter, provided that a convenient smoothing kernel is
used.
In their work, \citet{einasto_superclusters_2007} use an Epanechnikov
kernel, and assert that the best results in the identification method
are obtained when a kernel size of \mbox{8 $h^{-1}\,Mpc$} is used.
\citet{einasto_superclusters_2007} set the threshold parameter by
maximizing the number of large superclusters.
\citet{costaduarte_morphological_2010} apply the density field method
to the SDSS--DR7 and use two samples of structures, selected using
different overdensity thresholds.
The authors claim that there is no natural value for the threshold
density, and define a sample that maximizes the number of structures,
and a sample with the parameters tuned so that the largest
superclusters present an extension of \mbox{$\approx$ 120 h$^{-1}$
Mpc}.
The authors perform an analysis of the shapes of superclusters using
Minkowski functionals, finding that filamentary structures tend to be
richer, larger and more luminous than pancakes.
They also use a semi--analytic catalogue of galaxies derived from the
Millennium simulation to test the method and conclude that the
morphological classification is not biased by peculiar velocities.
\citet{liivamagi_sdssdr7_2010} use a similar approach, but perform the
smoothing with a B3-spline kernel of radius of \mbox{$8 h^{−1}\,Mpc$},
obtaining catalogues of superstructures with similar properties in the
SDSS--DR7 and in the Millennium simulation.

 \subsection{Present approach} \label{SS_OurMethod}

Given that not all large-scale structures are virialized at the
present time, the setting of identification parameters are subject to
certain degree of arbitrariness.
A physically motivated threshold on mass overdensity was explored
in $\Lambda CDM$ simulations by \citet{duenner_limits_2006}.
According to these authors, it is possible to define a criterion to
isolate, using three-dimensional data, overdensity regions enclosed by
a spherical shell that will evolve into virialized systems.
By the application of the spherical collapse model, the mean mass
density enclosed by the last bound shell of a structure must satisfy:
 
\begin{equation}
\frac{\bar{\rho}^{mass}_{shell}}{\bar{\rho}^{mass}_{bck}}=7.88,
\end{equation}

\noindent where $\bar{\rho}^{mass}_{shell}$ is the mean mass density
enclosed by the critical shell (the shell that maximizes the potential
energy), and $\bar{\rho}^{mass}_{bck}$ is the mean density of the
background.
In observational catalogues there is not an accurate estimation of
the mass density field.
However, given that at large scales the mass-luminosity ratio is nearly
constant, we can apply a similar criterion to the luminosity
map to derive structures with an appropriate mass overdensity.

In the next subsections we describe the identification method
implemented on the SDSS--DR7 spectroscopic data.
We first define the volume covered by the sample using a
three-dimensional mask and construct a luminosity density map with a
$1\,(h^{-1}\,Mpc)^3$ cell resolution.
We then use a percolation method based on the search of high density
peaks on the three-dimensional smoothed map.
These overdensities are the basis of the superstructure catalogue.

\begin{figure}
   \centering
   \includegraphics[width=0.5\textwidth]{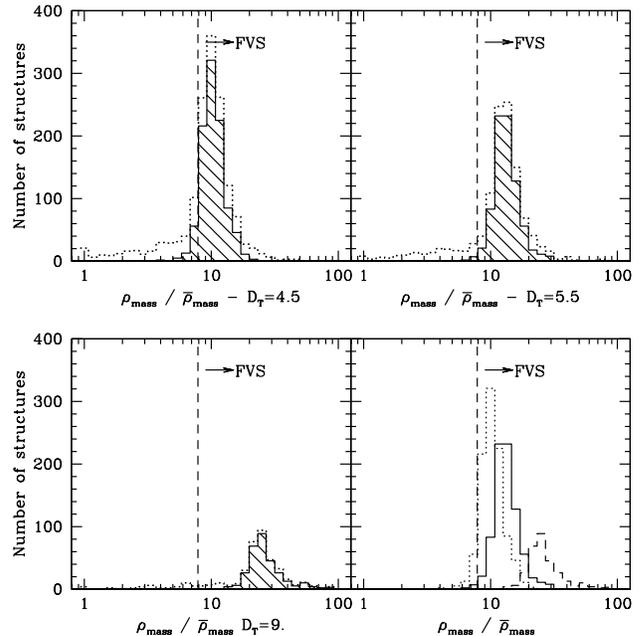}
   \caption{\change Distribution of the mass overdensity of the
   superstructures identified in the simulation box, for different
   values of the luminosity overdensity parameter ($D_T=4.5, 5.5$ and
   $9$, in the top left, top right and bottom left panels
   respectively).  The dashed vertical
   lines indicate the critical mass overdensity that
   a region must have to be virialized, according to
   \citet{duenner_limits_2006}.
   Dotted histograms show the distributions for the total sample of
   superstructures, and black continuous histograms show the 
   corresponding distributions for 
   structures with total luminosity greater than $10^{12}L_{\odot}$.
   Bottom right: comparison between the distributions of mass overdensity
   for the three values of $D_T$. \changeend}
   \label{fig:dlcs}
\end{figure}

\begin{figure}
   \centering 
   \includegraphics[width=0.5\textwidth]{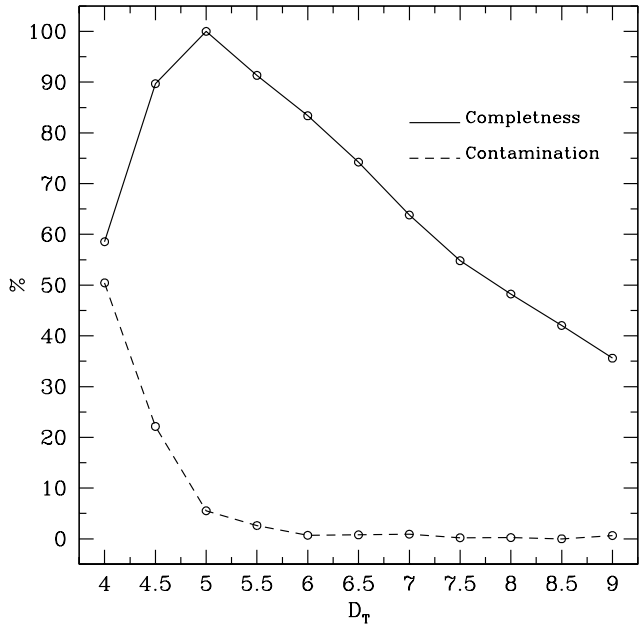}
   \caption{Completeness and contamination of the FVS catalogue as a function of
   $D_T$. \change The chosen value of $D_T=5.5$ for the compilation of the
   catalogue of FVS shows the best balance between
   low contamination and high completeness.\changeend}
   \label{fig:cyc} 
\end{figure}  


 \subsubsection{Luminosity Density Map} \label{SS_kernel}

We construct smooth luminosity density maps from our complete samples
of galaxies, in the region defined by a three-dimensional mask.
This mask, which represents an approximation of the geometry of the
SDSS--DR7 galaxy catalogue in both angular coordinates and redshift,
is built using cubic cells of \mbox{$1\, h^{-1}\,Mpc$} side.
We start from a pixelized representation of the central region within
the catalogue footprint area.
To this end we use an angular mask obtained using the software 
\textsc{HEALPIX}
\citep{gorski_healpix_2005} with a resolution parameter $N_{side}=512$
that splits the whole sphere in $3\,145\,728$ equal area pixels.  A
cubic cell is part of the 3D mask if at least a fraction of its area
includes part of the solid angle subtended by the angular mask, and
its radial position lies within the redshift range defined for a given
sample.
\change
We define the volume of a cell within the three--dimensional mask as
$V'_{cell} = V \times f_{cell}$, where $V=(1h^{-1}Mpc)^3$ and $f_{cell}$
measures the fraction of the volume $V$ into the mask.
We compute the fraction $f_{cell}$ by implementing a Monte Carlo
procedure.
Accordingly, cells that are completely contained within the geometry
of the catalogue have $f_{cell}=1$, while external cells have
$f_{cell}=0$.
The factor $f_{cell}$ allow us to correct for border effects, in our subsequent
analysis we will use only cells with $f_{cell} > 0.5$ i.e., only cells
with at least 50\% of their volume inside the mask.
\changeend
The continuous luminosity density map can be constructed by smoothing
the galaxy distribution within the 3D mask.
A standard procedure to accomplish this consists in using a kernel
function to convolve the discrete positions of galaxies
and spread their luminosity.
The resulting density field is then represented at a resolution given
by the cell size, and corrected by the weight factor $f_{cell}$.
The result of the smoothing depends on the shape and size of the
chosen kernel function.
Following previous analysis in the literature
\citep{einasto_richest_2007,costaduarte_morphological_2010}, we use an
Epanechnikov kernel of \mbox{$r_0=8\,h^{-1}$Mpc} size, which gives the
contribution at position ${\bf r}$ from a source that is located at
${\bf R}$:
              
\begin{equation}
\label{kernel}
k({\bf r}-{\bf R})=\frac{3}{4r_{0}}\left[1-\left(\frac{|{\bf r}-{\bf R}|}{r_{0}}\right)^{2}\right].
\end{equation}

\noindent
An Epanechnikov kernel is more suitable for
this analysis since its shape resembles that of a Gaussian, but it
avoids excessive smoothing.
%
The luminosity density estimate within a cell is then:
\change
\begin{equation}
\rho_{cell}=L_{cell}/V'_{cell},
\end{equation}

\noindent where 
$V'_{cell}=V\times f_{cell}$ is the volume of the cell and
\changeend
$L_{cell}$ is the sum of the contributions to the
luminosity $L_{glx}$ from nearby galaxies:

\begin{equation}
L_{cell}=\sum_i L_{glx}^i\int_{cell} \; k({\bf r}-{\bf R_i})d{\bf r}.  
\end{equation}

\noindent A slice of the map is shown in \fref{f_Ldensmap}, where the
large scale structure of the luminosity distribution can be
appreciated. 
Concentrations of galaxies such as rich groups or clusters are
characterized by luminosity density peaks, and are surrounded by low
density regions. 


 \subsection{Identification of structures} \label{SS_iden}

We search for large isolated regions above a given density threshold;
these are candidates to FVS.
The luminosity density, however, depends on the limiting magnitude
characterizing a given volume limited sample, which determines the
number and luminosity of galaxies contributing to the overall density
estimate.
For example, deeper volume limited samples include only brighter
galaxies, thus the luminosities of the cells will be systematically
lower as the upper redshift limit of the sample increases.
We use the luminosity overdensity instead of the total luminosity density
value to characterize the peaks in the luminosity distribution, but
still, since we study three samples with different limiting redshifts,
a correction factor must be implemented in order to make the results
as comparable as possible.
This factor can be obtained by assuming a universal luminosity
function for galaxies $\Phi(L)$, which allow us to account for the
luminosity below the limiting value characterizing a given sample.
Since the underestimation of total luminosity depends only on the
absolute magnitude cut $M_{lim}$ of the sample, an homogeneous
correction factor $F$ for the entire luminosity density map can be
defined as:

\begin{equation}
F=\frac{\int_{0}^{\infty}L\Phi (L)dL}{\int_{M_{lim}}^{\infty}L\Phi (L)dL}.
\label{F_correctionL}
\end{equation}

\noindent
We adopted the luminosity function presented by
\citet{blanton_galaxy_2003}.
This correction ranges from $~2$ to $\lesssim 6$ for samples with
limiting absolute magnitudes down to $M_r=-21$.
In \tref{T_samples} the correction factor $F$ and the
corrected mean luminosity density are listed for each sample.

We define superstructures by linking overdense cells using a Friends-of-Friends 
algorithm that connects overdense cells sharing at least one
common vertex or side.
To this end, we use a threshold luminosity overdensity criterion
$\rho_{lum-cell}\geq D_T \, \bar{\rho}_{lum}$.
Since the properties of the catalogue of superstructures may be
affected by the adopted value of
\change
 the luminosity overdensity parameter $D_T$,
\changeend it is fundamental to study
this issue into further detail.

 
 \subsection{Calibrating the method with numerical simulations}\label{SS_choiceD}

We use numerical simulations to test the ability of a
luminosity--based algorithm to derive structures that will evolve into
virialized systems in the future.
We have also used the simulations to analyse whether superstructures
identified in redshift--space with total luminosities above
$L=10^{12}\, L_{\odot}$, correspond to systems in real--space above
this threshold.
This is an important issue to be checked since the observations
provide redshift--space data whereas the FVS criterion requires
real--space information.
In \fref{fig:dlcs} we show the distributions of mass overdensity 
(only known in the simulations) for the sample of structures that
result from the identification algorithm for three different values of
$D_T$, and the corresponding distributions for superstructures with a
luminosity above the threshold \mbox{$10.^{12}$ L$_{\odot}$}.
This adopted luminosity threshold is approximately the maximum
observed value for a rich cluster of galaxies; by imposing this
luminosity cut--off we exclude individual clusters from our FVS
catalogue.  
The bottom right panel of \fref{fig:dlcs} shows the mass overdensity of
superstructures with the luminosity cut--off for three values in
luminosity overdensity parameter.
The vertical dashed line in this figure shows the minimum mass overdensity
 for a superstructure to be virialized in the future according
to \citet{duenner_limits_2006}.
As can be seen, the contamination in the catalogue depends on the
choice of the luminosity overdensity parameter.
In order to select the most convenient value of this parameter, we
explored the contamination and completeness of superstructures for
different values of $D_T$.
We define the completeness as the ratio of the number of
superstructures resulting from a given value of $D_T$ and 
 maximum number of superstructures obtained within the explored
$D_T$ range.
Similarly, the contamination is defined as the fraction
of identified superstructures that do not satisfy the future
virialized criteria. 
In \fref{fig:cyc} we show the completeness and contamination
parameters for the range $4<D_T<9$. By inspection of this figure, we
chose a luminosity overdensity threshold $D_T=5.5$.
This adopted value provides a suitable compromise of high completeness
and low contamination.


 \subsection{Redshift--space vs. real--space analysis}\label{SS_ZvsR}

\begin{figure*}
\centering
\hfill%
\subfigure[ Luminosity distribution of FVS]{%
\label{fig:Lmockvsbox}%
\includegraphics[width=.5\textwidth]{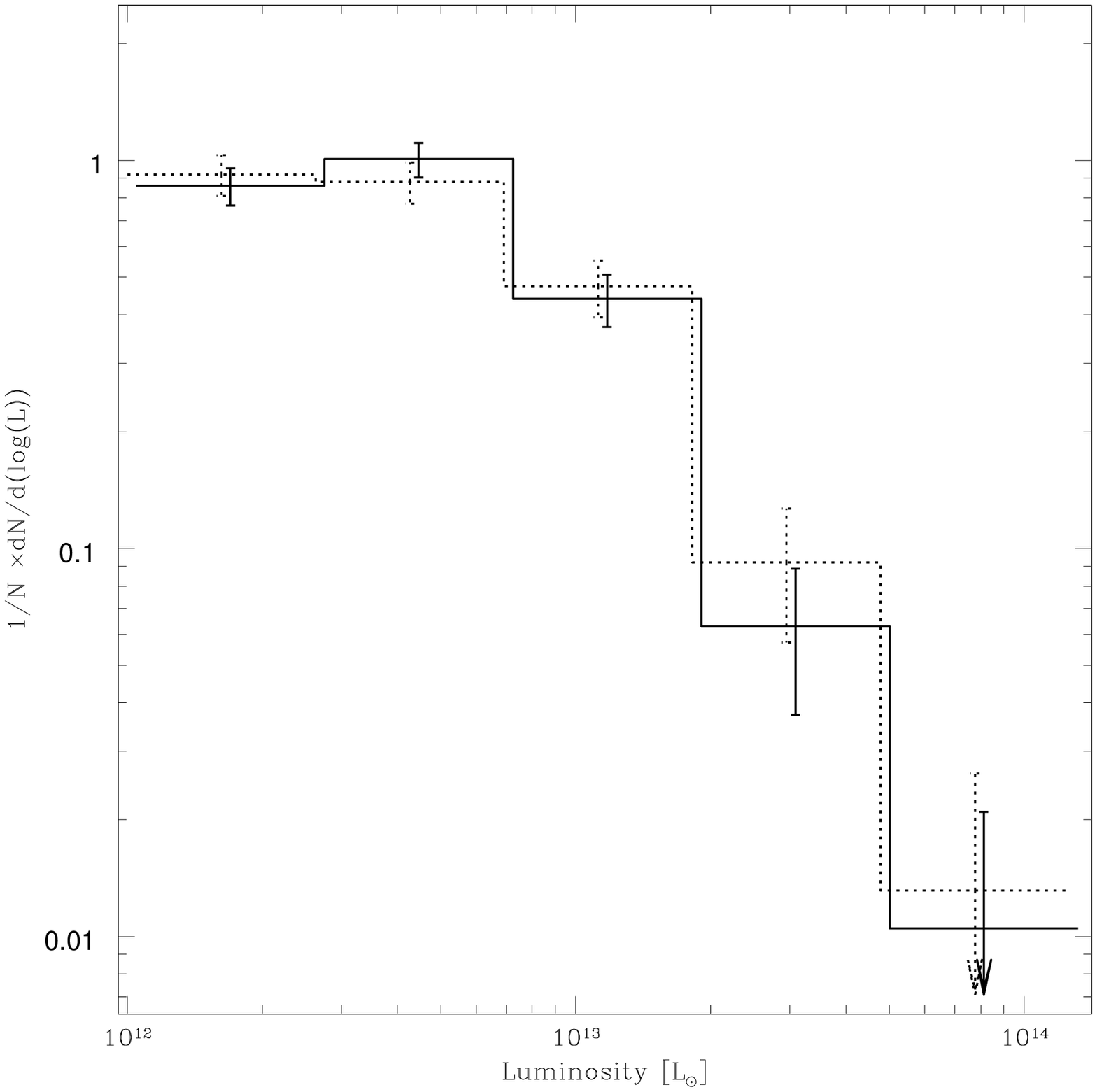}}~\hfill
\subfigure[ Volume distribution of FVS]{%
\label{fig:Vmockvsbox}%
\includegraphics[width=.5\textwidth]{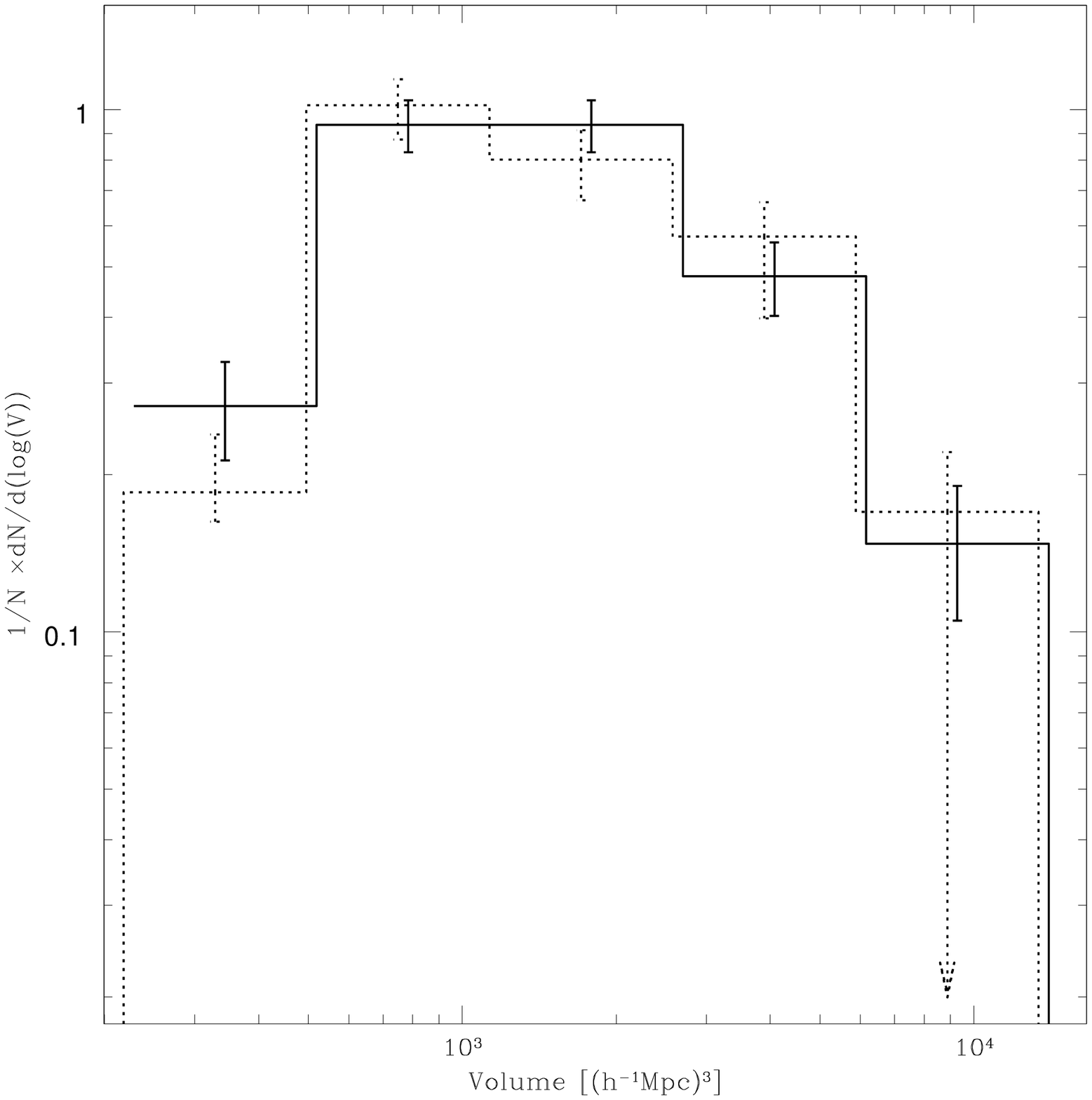}}\hfill~
\caption{Histograms representing the luminosity and volume distributions of
   Future Virialized Structures.  Solid lines correspond to the
   sample of FVS identified in sample $M_{Zsp}$ (redshift--space mock, see \sref{S_data}), 
   and dotted lines correspond to FVS identified in sample
   $M_{Rsp}$ (real--space mock).
}
\end{figure*}

In this subsection we analyse the effects of redshift--space
distortions induced by peculiar velocities on the identification of
FVS, using the mock catalogue.
We considered the samples $M_{Rsp}$ and $M_{Zsp}$ (\tref{T_samples})
which use the real--space and redshift--space position of mock
galaxies, respectively, and are
defined using the same redshift limits and limiting absolute magnitude
than sample S2.
This allows to study FVS identified in real--space and compare
them with those selected in redshift--space.
We have computed the distributions of luminosity and volume derived
for the FVS in these mock samples.
A good agreement can be appreciated between the luminosity
distributions (\fref{fig:Lmockvsbox}), indicating that the
identification procedure delivers reliable estimations of the total
luminosity.
It can also be inferred from the volume distributions, shown in
\fref{fig:Vmockvsbox}, that the volume of FVS are not strongly
affected by peculiar velocities.

We also analysed the correspondence between structures in real and
redshift--space in order to estimate the contamination and
completeness of the FVS catalogue.
The first point we notice is the lower number of FVS identified in the
redshift--space sample in comparison to the real--space sample, in the
same volume.
About 20 per cent of FVS identified in real--space are lost when analyzing
redshift--space data.
This is the major drawback of redshift--space FVS identification.
On the other hand, only 2 per cent of FVS identified in redshift--space are
not real--space FVS. 
This indicates that the identification of FVS in observational
catalogues is not likely to include fake FVS structures.
We also found that approximately 3 per cent of FVS in the mock catalogue are
associated to more than a single FVS in real--space.
A similar percentage of real--space FVS are identified as multiple FVS
in redshift--space.


\section{Catalogue of FVS}\label{S_catalog}

Taking into account the results of the previous sections,we have
adopted a luminosity overdensity threshold of $D_T=5.5$ and
in order to avoid the inclusion of spurious systems, a lower
luminosity limit  $L_{str}>10^{12}L_{\odot}$.
In \tref{T_results}
we summarize some characteristics of the identified FVS.

\begin{table}
\centering
\begin{tabular}{@{\extracolsep{\fill}}ccccc}
\hline 
Sample & $N_{FVS}$ & $F_{vol} $  & $F_{lum} $ & $glxs_{inFVS}$ \\
\hline
S1     & 67  & 1.08\%  & 10.85\% &  9707  \\
S2     & 150 & 1.26\%  & 13.54\% &  11394 \\
S3     & 412 & 1.66\%  & 20.61\% &  11682 \\
\\
$M_{Rsp}$& 227 & 1.62\%  & 18.87\% & 19265 \\
$M_{Zsp}$& 181 & 1.35\%  & 15.14\% & 15368 \\
\hline
\end{tabular}
\caption{Main results obtained for the samples of identified FVS.
For each sample, we show the number of future virialized structures
$N_{FVS}$, the percentage of volume occupied by FVS $F_{vol}$,
the percentage of luminosity of galaxies within FVS $F_{lum}$ and the
total number of galaxies within FVS $glxs_{inFVS}$.}
\label{T_results}
\end{table}

Once the superstructures have been identified, it is important to
assess their fundamental properties such as total volume, number of
galaxies above a given luminosity, the total luminosity, shape
parameters, etc.
Given the small volume of sample S1, and the large luminosity
correction factor in sample S3, we provide an analysis of sample S2 in
what follows.
%


\subsection{Analysis of the FVS catalogue (sample S2)} \label{SS_cat:vol}

\begin{figure*}
\centering
\hfill%
\subfigure[ Luminosity distribution of FVS]{%
\label{fig:Hlum}%
\includegraphics[width=.5\textwidth]{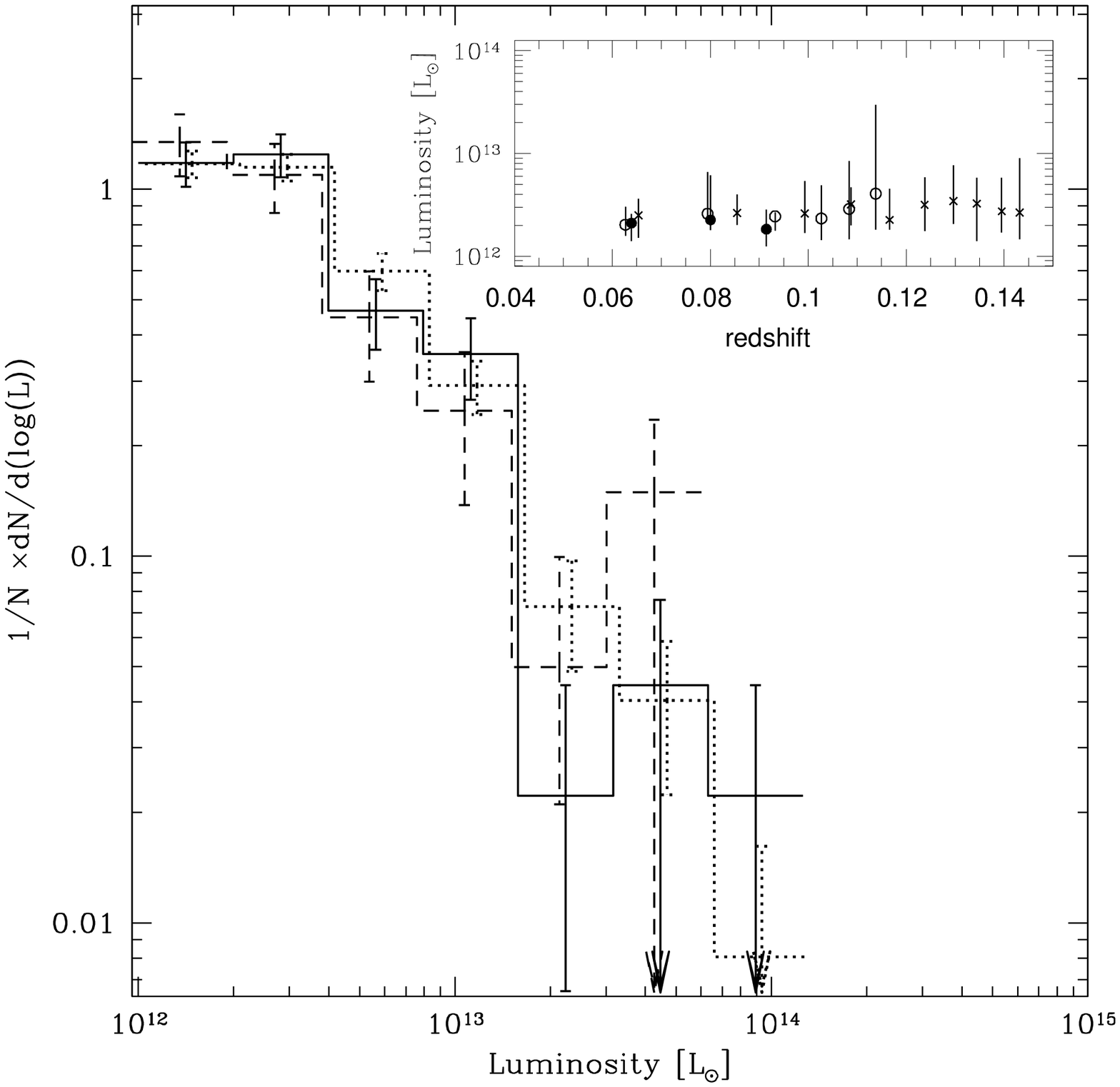}}~\hfill
\subfigure[ Volume distribution of FVS]{%
\label{fig:Hvol}%
\includegraphics[width=.5\textwidth]{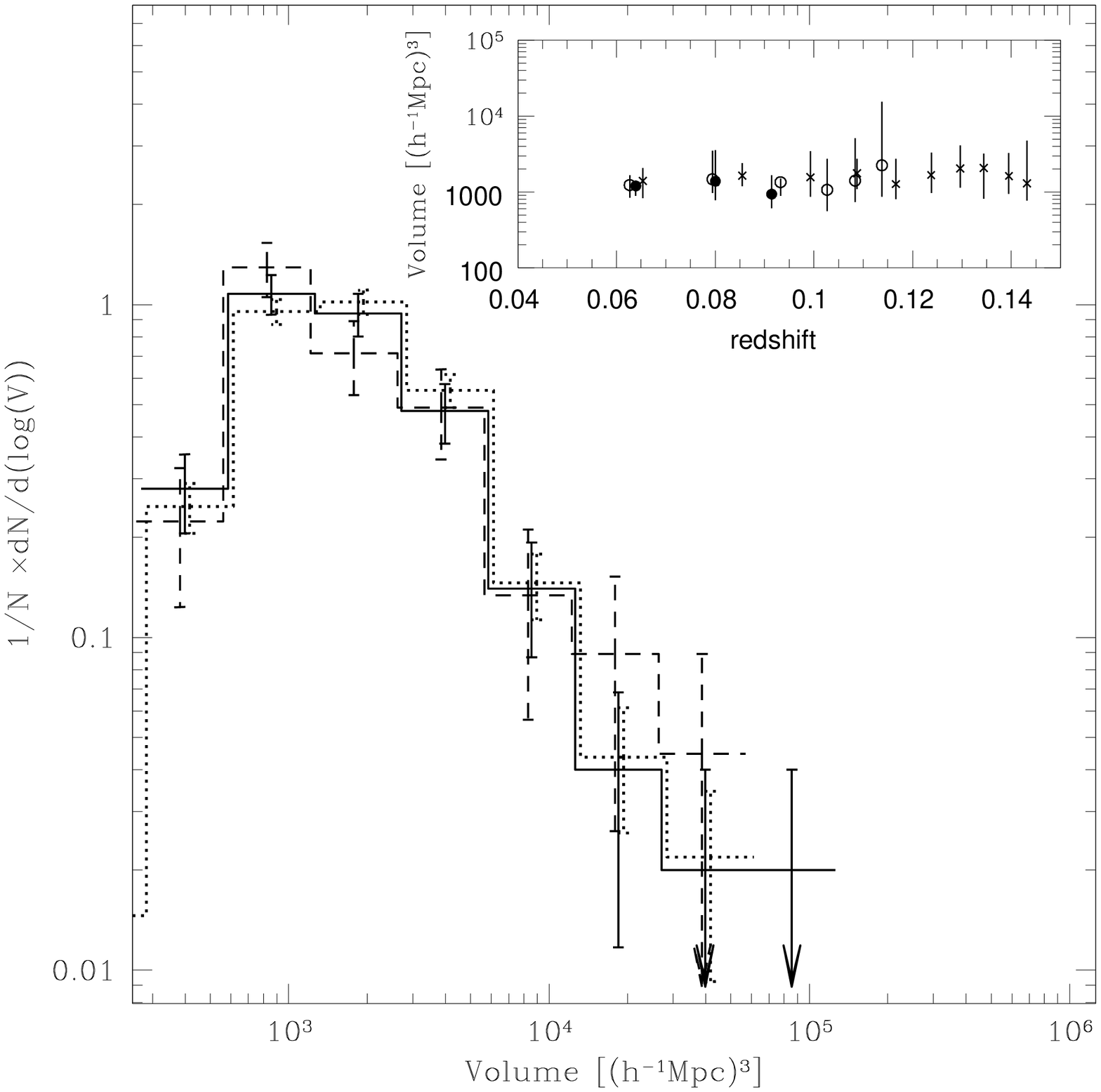}}\hfill~

\caption{ Luminosity and volume distributions of structures for
samples S1 (dashed line), S2 (solid line) and S3 (dotted
line).  The insets show the average volume and luminosity as a
function of the redshift, in redshift bins with equal numbers of
superstructures, with filled circles, empty circles and cruxes,
respectively for samples S1, S2 and S3.  Error bars on the histograms
indicate Poisson uncertainty.  In the inset box, the uncertainty bars
correspond to the 25 per cent and 75 per cent percentiles.}

\end{figure*}

In \fref{fig:Hlum} we show the distribution of luminosities of the
superstructures we obtain for the SDSS--DR7.
FVS luminosities vary between $10^{12}L_{\odot}$ and $\sim
10^{14}L_{\odot}$, in agreement with \citet{einasto_luminous_2006} and
\citet{costaduarte_morphological_2010}.
However, \citet{einasto_luminous_2006} find a lack of luminous
superclusters in numerical simulations compared to superclusters
identified in observational catalogues.
In a forthcoming paper (Luparello et al., in preparation) we will
analyse this issue in more detail.

Since the superstructures are obtained from a discrete density map,
the volume can be directly calculated as the sum of the volumes of all
the cells that belong to a given system.
As all cells have the same volume, equal to \mbox{1 $(h^{-1}Mpc)^3$},
the number of cells that form a structure is directly proportional to
the volume.
As it can be appreciated in \fref{fig:Hvol}, the volume of the
superstructures ranges between  \mbox{$10^2 (h^{-1} Mpc)^3$} and
\mbox{$10^5 (h^{-1} Mpc)^3$}.
We also compute the fraction of the total volume that is
in FVS, finding that FVS represents only 1.26 per cent of the
volume covered by the sample S2.
\citet{einasto_comparison_simulations} found that the volume covered
by superclusters in the 2dFGRS represents 3.2 per cent of the Northern region
of the catalogue and 3.5 per cent in the Southern region. 
Our lower percentages can be explained by the fact that we use a
higher value of $D_T$ and impose a limit of at least
$10^{12}L_{\odot}$ in the luminosity of the superstructures.

\begin{figure} 
   \centering
   \includegraphics[width=0.5\textwidth]{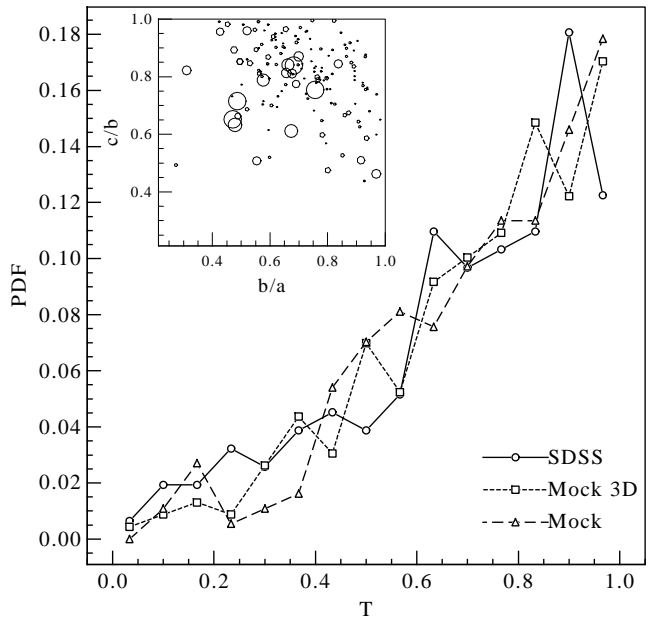} 
   \caption{
   Probability density distribution estimates of the shape indicator
   parameter $T$ (defined in section \sref{SS_cat:vol}).  The
   distributions correspond to FVS from the SDSS galaxy catalogue
   (solid line), the real--space mock catalogue (short dashed line) and the
   redshift--space mock catalogue (long dashed line).
   The inset shows a scatter-plot of the semi--axis ratios $c/b$ and $b/a$
   that characterize the shapes of the FVS.  The sizes of circles are
   proportional to the number of galaxies contained in each FVS.}
   \label{fig:shapes_sdss} 
\end{figure}

Previous analyses of numerical simulations have shown that the shapes
of halos are consistent with an increasing elongation as the mass
increases. This is also detected in observational data
\citep{paz_shapes_2006}, where the most massive structures tend to be
prolate.
We notice that the shapes of FVS are somewhat irregular since several
of these systems are composed by a few filaments or pancakes of
irregular shapes, joined by one or two vertexes.
Nevertheless, in order to provide a rough characterization of FVS
shapes, we use the usually adopted ellipsoidal model to study the
distribution of axes ratios.
We have computed the moment of inertia $I_{ij} = \sum_{gxs}
(x_i^{gx}-C_i^{FVS})*(x_j^{gx}-C_j^{FVS})$, where $x_i^{gx}$ is the
i-th coordinate (i.e., the $x$, $y$, or $z$ axis) of the member
galaxies, and $C_i^{FVS}$ are the coordinates of the geometrical
centre of each FVS.
The diagonalization of this matrix gives the values of the three axis
of the ellipsoid with the same moment of inertia , $a$, $b$ and $c$.
In the inset of \fref{fig:shapes_sdss} we show the results of the
axial ratios of FVS in an $c/b$ vs $b/a$ diagram.
As can be seen in this figure, neither spherical ($a \sim b \sim c$)
nor planar ($b \sim a > c$) structures are common; there is a
preference for prolate shaped FVS structures.
We use the triaxiality parameter $T=(1-(b/a)^2)/(1-(c/a)^2)$  adopted
by \citet{bahcall_2007}, who analysed supercluster shapes in a
$\Lambda CDM$ model simulation. 
This parameter takes values between 0 and 1; $T \sim 0$ indicates
oblate structures, while $T \sim 1$ indicates prolate structures.
Intermediate values of $T$ represent triaxial systems, with
triaxiality increasing (not linearly) with $T$.
We show in \fref{fig:shapes_sdss} the distribution of $T$ for samples
S2, $M_{Zsp}$ and $M_{Rsp}$.
As it can be seen from this figure, the distribution of shapes is not
strongly affected by peculiar velocities. 
Also, it can be appreciated that the observed SDSS FVS shapes are
consistent with those of $\Lambda CDM$ structures.
The mean value of this parameter is $T=0.75\pm0.23$ for the main
sample of FVS of the SDSS, and $T=0.77\pm0.23$ for the sample of the
real--space simulated catalogue. 
This indicates that these structures are mainly triaxial systems, and
the distribution of $T$ shows the predominance of prolate systems over
oblate systems, in agreement with
\citet{einasto_comparison_simulations}.
The mean values of $T$ reported by \citet{bahcall_2007} range between
$T=0.65$ and $T=0.69$ depending on the linking length, in agreement
with our results.
\citet{costaduarte_morphological_2010} present evidence of a trend
between supercluster luminosities and shapes, where filaments are on
average more luminous than pancakes.
We also explored the dependence of the $T$ parameter on FVS luminosity
and volume, finding a similar tendency. Larger values of $T$
correspond to more luminous, larger, FVS. 
%

\subsection{Testing the reliability of the method} \label{SS_cat:tests}

\subsubsection{Redshift dependence} \label{SS_zdependence}

We have also searched for a possible redshift dependence of our
results; notice that given the shallow depth of our samples this
constitutes a test of our method rather than a search for true time
evolution.
To this aim, we use the samples S1 and S3, defined in section
\ref{S_data}.
The similarity of the three histograms in
\fref{fig:Hlum} and \fref{fig:Hvol} corresponding to these different
subsamples shows a lack of redshift dependence.
This is also seen in the insets of \fref{fig:Hlum} and
\fref{fig:Hvol}, in the dependence of the median luminosity and volume
with redshift.
As it can be seen, there is no significant dependence of the median
luminosities and volumes with redshift, indicating the reliability of
the method and its lack of a redshift bias.

\subsubsection{Luminosity cut--off dependence} \label{SS_Mdependence}

\begin{figure*}
\centering
\hfill%
\subfigure[ Luminosity distribution of FVS]{%
\label{fig:Hlum-Mc}%
\includegraphics[width=.5\textwidth]{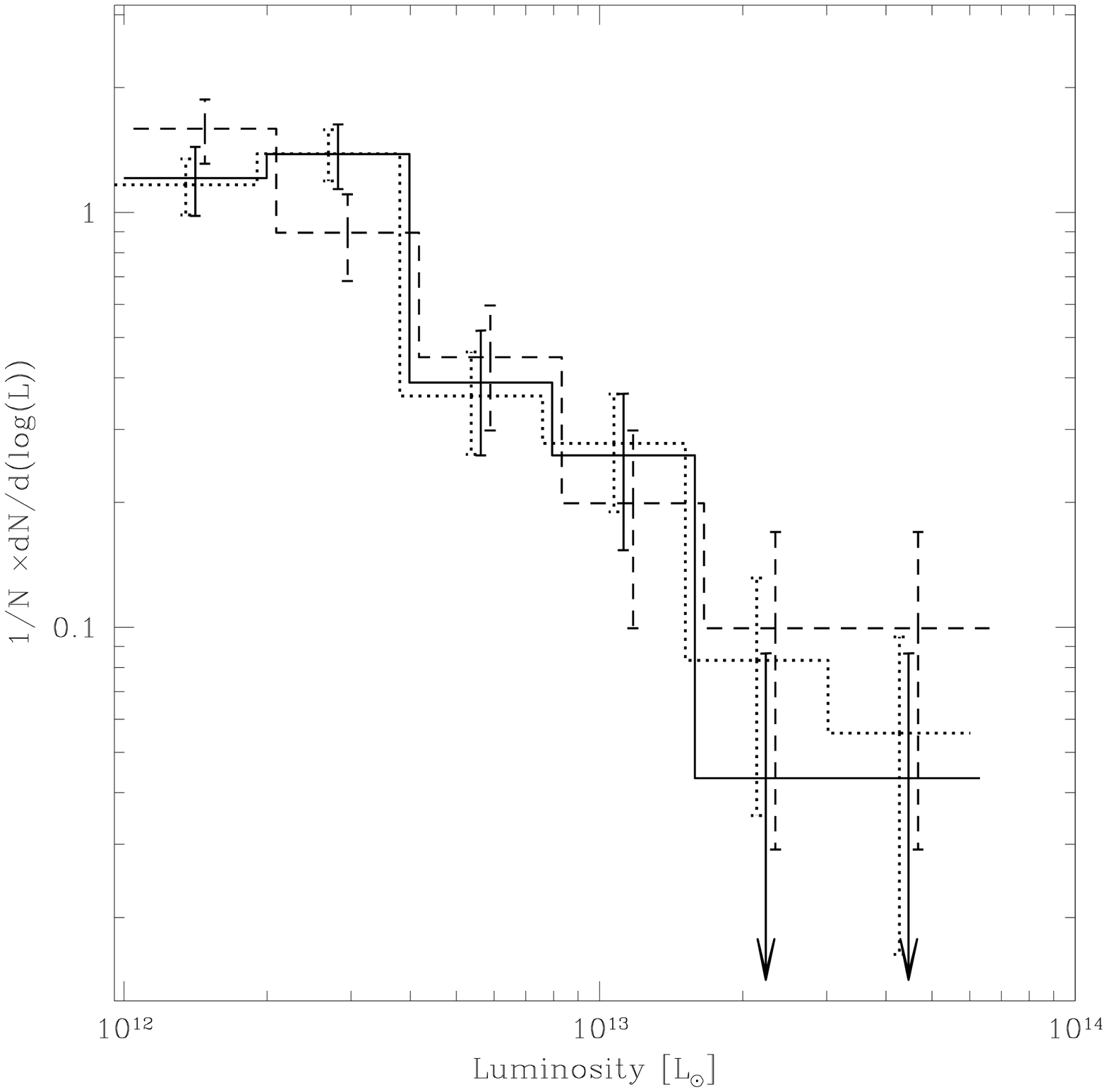}}~\hfill
\subfigure[ Volume distribution of FVS]{%
\label{fig:Hvol-Mc}%
\includegraphics[width=.5\textwidth]{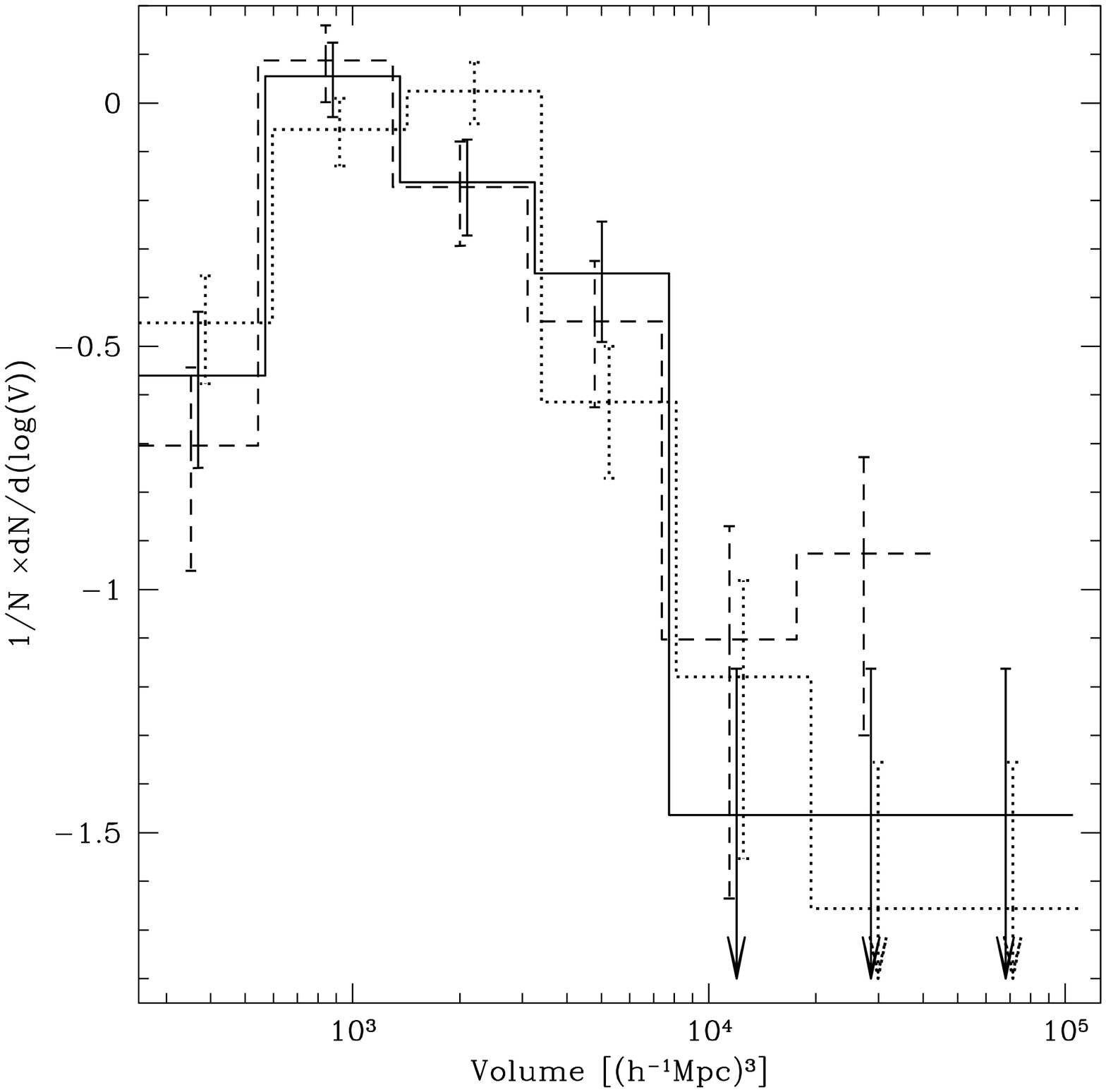}}\hfill~
\caption{ Luminosity and volume distributions of structures for samples S1
   (dashed line), S2c (solid line) and S3c (dotted line).
   These three samples are limited to $z\le 0.1$, but comprise 
   galaxies with different absolute limiting magnitudes (see
   \tref{T_samples}).
   Error bars are Poisson uncertainties within each bin. 
}
\end{figure*}

With the aim of analyzing the effects of using tracers of different
luminosities, we applied three magnitude cuts, corresponding to those
of samples S2 and S3, to galaxies in the redshift range $0.04\leq z
\leq 0.10$.
This defines two new samples: S2c and S3c, as described in
\tref{T_samples}.
In figures \ref{fig:Hlum-Mc} and \ref{fig:Hvol-Mc} we show the
distribution of luminosities and volumes of FVS selected from the S1,
S2c and S3c samples.
We recall that we have applied the corresponding completeness
luminosity factor $F$ (Eq. \ref{F_correctionL}) to the luminosities of FVS.
As can be seen in these figures, there is a very good agreement
between the FVS derived from these different galaxy tracers.
These tests give confidence that the use of volume limited samples of
high luminosity galaxies do not change significantly the results, with
the advantage that they can trace larger volumes.

\section{Correspondence of known Superclusters and FVS} \label{S_SCSvsFBS}

As mentioned in \sref{S_method}, superclusters have been previously
identified either as luminosity density enhancements, or as systems of
galaxy clusters.
None of the criteria ensure that the structures will evolve into
gravitationally virialized structures.

\begin{figure*}
   \label{F_celebrities} \centering \hfill%
   \subfigure[Sloan Great Wall]{\label{fig:SGG}%
   \includegraphics[width=.5\textwidth]{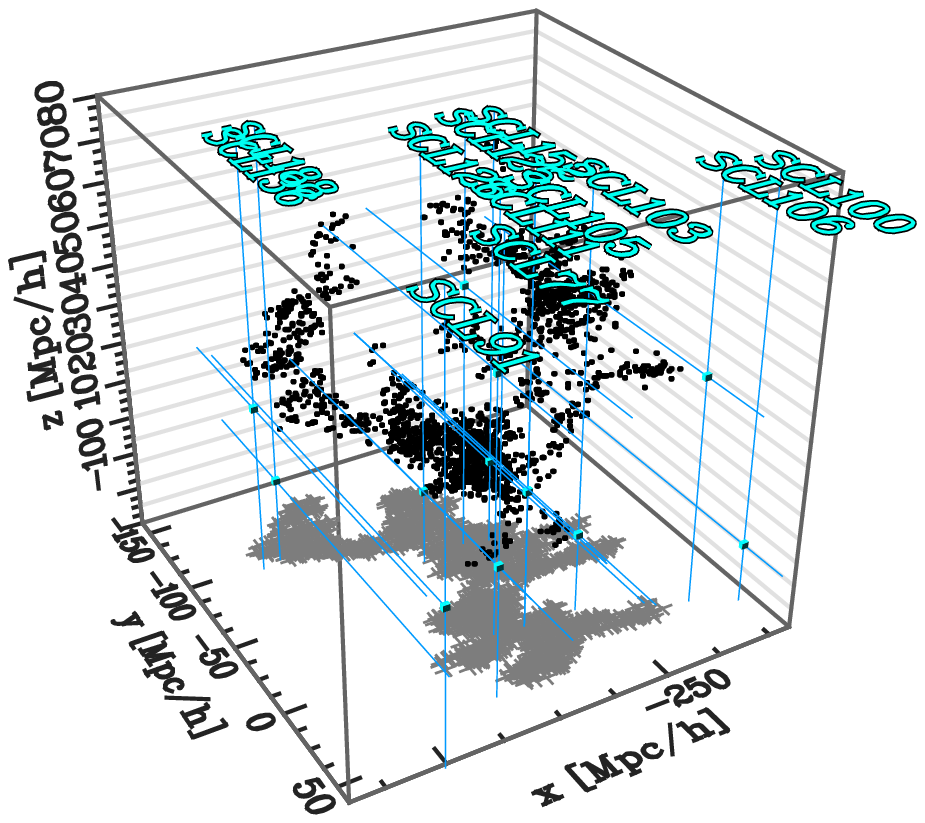}}~\hfill
   \subfigure[Ursa Major supercluster]{\label{fig:UM}%
   \includegraphics[width=.5\textwidth]{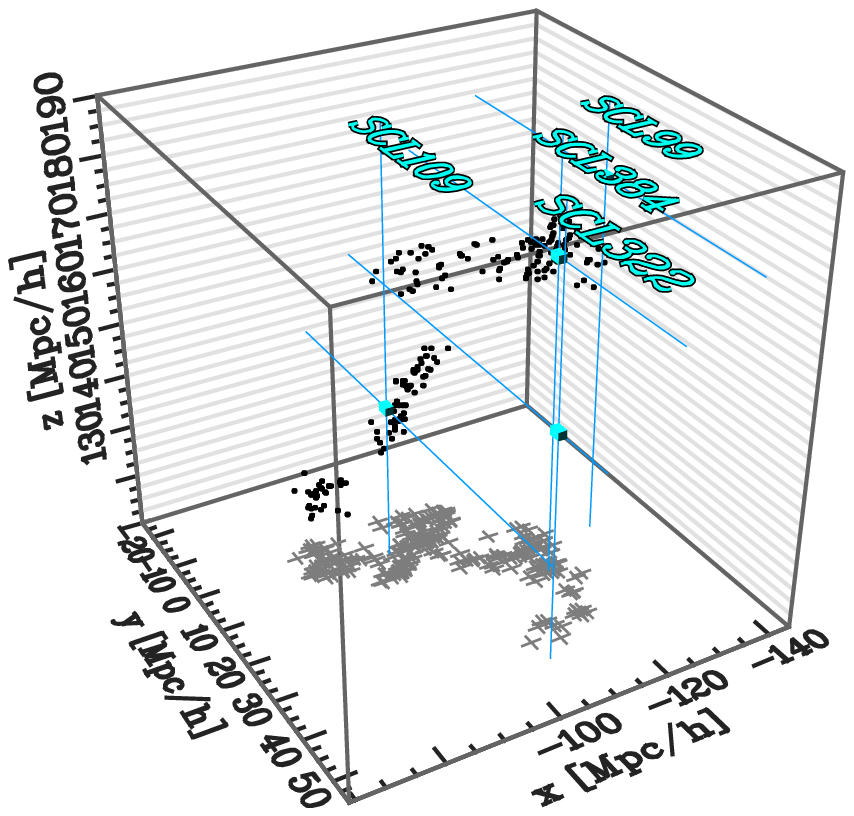}}\hfill~
   \subfigure[Corona Borealis supercluster]{\label{fig:CB}%
   \includegraphics[width=.5\textwidth]{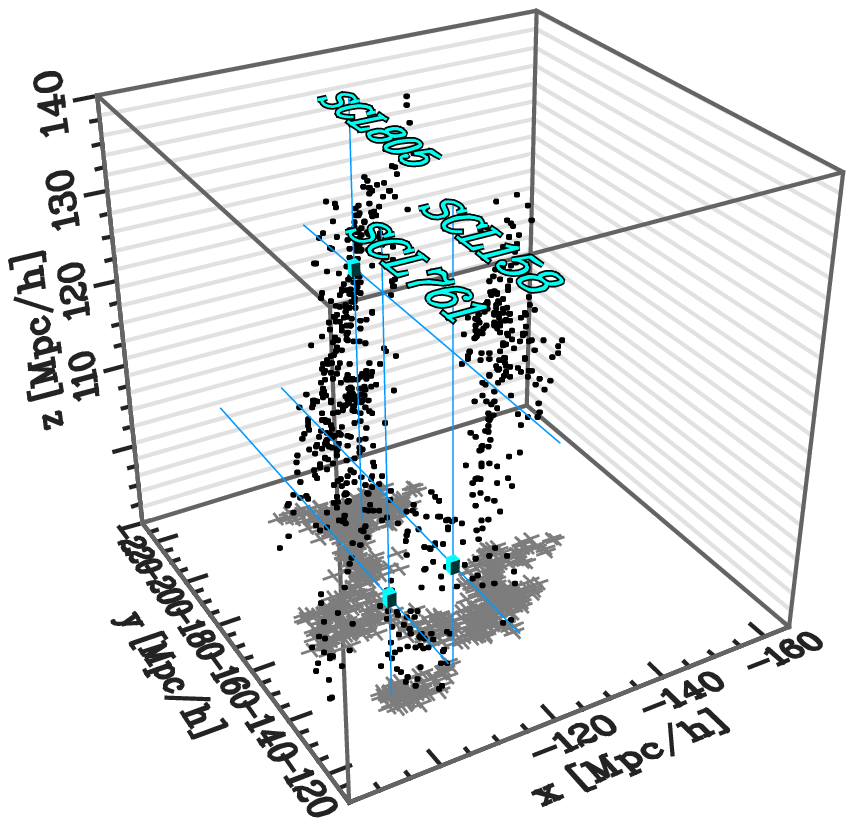}}~\hfill 
   \subfigure[Bootes supercluster]{\label{fig:BOO}%
   \includegraphics[width=.5\textwidth]{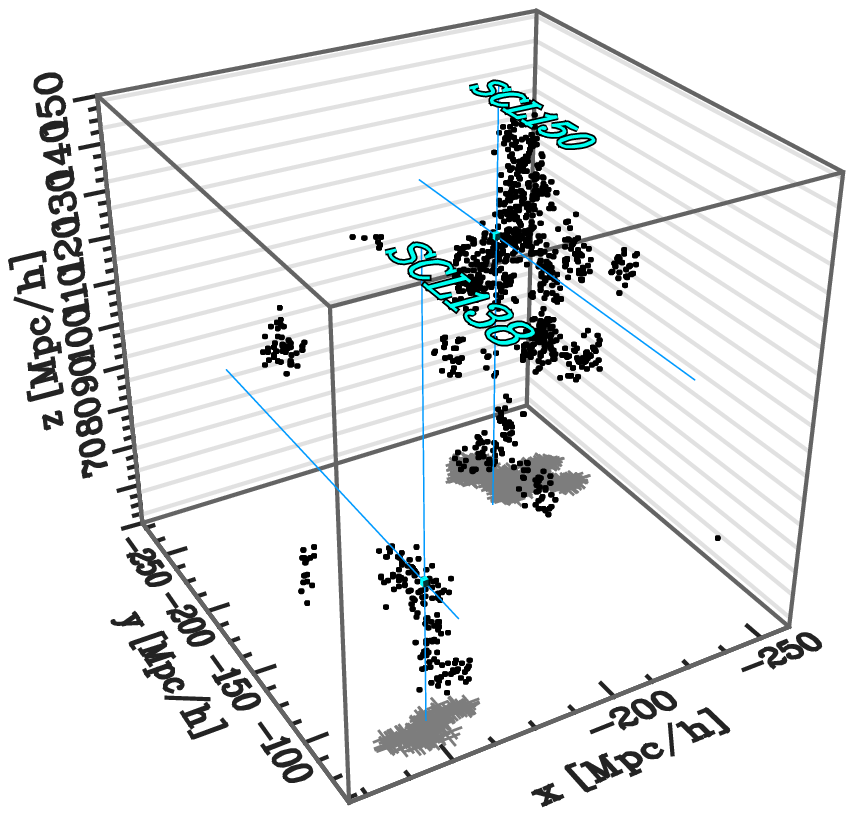}}~\hfill 
   \caption{%
   Spatial distributions of galaxies within 
   Future Virialized
   Systems associated to known superclusters (filled circles).
   The positions are shown in a cartesian system associated to the equatorial
   coordinate system (x-axis pointing to the vernal point $\gamma$ and z-axis pointing to
   north equatorial pole); the line indicates the observer direction.
   We also show the centres of superclusters in the catalogues of 
   \citet{einasto_optical_2001}, \citet{einasto_superclusters_2007}
   and \citet{einasto_supercluster_2006}.
   Galaxies within nearby FVS are indicated with \change black dots.
   In the Sloan Great Wall region (a), we identify a single FVS associated
   with SCL111, SCL126 and SCL136, indicating that these superclusters may
   evolve into a single virialized system. 
   The Ursa Major Supercluster, in (b), consists in three large filaments \citep{kopylova_detailed_2006}.
   However, we find a different FVS associated to each filament, suggesting
   that the filaments would evolve into separate structures.
   In (c), the Corona Borealis Supercluster SCL158, is part of a larger FVS, that also
   includes SCL805 and SCL761. So, these three systems could merge to form a single
   virialized structure in the future.
   Bootes and BootesA Superclusters in (d) have one-to-one correspondence with FVS, so
   each system would evolve into a future virialized structure.\changeend
   }
\end{figure*}

Therefore, it cannot be ensured that all known superclusters will be
'island universes' in the future.
Recent supercluster catalogues were compiled from the 2dFGRS
\citep{einasto_superclusters_2007}, the SDSS--DR4
\citep{einasto_supercluster_2006} and the SDSS--DR7
\citep{costaduarte_morphological_2010}.
Since some of these catalogues cover only partially the volume of the
central zone of SDSS--DR7, we also used the supercluster
identification of \citet{einasto_optical_2001}, where the
identification is based on X--ray selected and Abell clusters.
Combining data from these catalogues we can correlate our results with
previous identifications and explore which known superclusters will
also be future virialized structures, at least within the $\Lambda
CDM$ scenario.
In \fref{F_celebrities} we show the FVS associated to the Sloan Great
Wall, Ursa Major, Corona Borealis and Bootes superclusters.
This figure shows the galaxies belonging to FVS associated to each of
the above mentioned superclusters, and also the galaxies in
neighboring FVS.
For comparison, the locations of the centres of superclusters in the
same region obtained from the literature are presented.

\textbf{Sloan Great Wall}.
There are several identified structures in the region associated to
the Sloan Great Wall.  
\citet{einasto_optical_2001} claim that the structures in their
catalogue named SCL111 (\textit{a.k.a.} Virgo--Coma) and SCL126 are
part of the Sloan Great Wall.  
Of these superclusters, SCL126 is the richest in the Sloan Great Wall.
\citet{einasto_sloan_2010} find that the structure previously
identified as SCL136 in \citet{einasto_optical_2001} is in fact a part
of the SCL126  
\change
,and SCL111 comprises three concentrations of rich clusters connected by
filaments of galaxies. 
\citet{einasto_sloan_2010} also find the presence of substructures, indicating
possible mergers and infall in different superclusters belonging to the 
Sloan Great Wall region, suggesting that the cores of these superclusters
are not virialized and still assembling.
\changeend
We find that, according to the catalogue of FVS, SCL111 and SCL126 may
evolve into a single virialized structure, while the neighboring 
SCL91 could remain isolated.  
The structure SCL100 \citet{einasto_optical_2001}, associated with the
Leo A supercluster and located near the Sloan Great Wall, has no FVS
that it can be directly associated with.
We find several FVS surrounding the centre of SCL100 thus suggesting
that this supercluster could be disrupted in the future.

\textbf{Ursa Major Supercluster}.
Ursa Major is a nearby \mbox{$z\simeq 0.06$} and relatively isolated
supercluster.  
It has been found to be composed of three large filaments with mean
redshifts z=0.051, 0.060 and 0.071 \citep{kopylova_detailed_2006}.  
\citet{einasto_optical_2001} associate  Ursa Major to the SLC109
superstructure in their catalogue, and present its geometrical centre
located at \mbox{$\alpha=177.1$ deg}, \mbox{$\delta=+55.0$ deg} and
$z=0.06$. 
\citet{einasto_supercluster_2006} identifies the more distant
filament as another individual supercluster (SCL384).
We found four different FVS with filamentary structure associated to
this supercluster.
This indicates that the filaments would eventually evolve into
separate structures.

\textbf{Corona Borealis Supercluster}. 
The Corona Borealis supercluster (SCL158 in
\citet{einasto_optical_2001}) is a prominent example of a supercluster
in the northern sky.  
This structure comprises $\simeq 500$ galaxies at $z=0.07$, with a
nearly spherical morphology.
We find that Corona Borealis is part of a larger FVS, which also
includes the superclusters SCL805 and SCL761 identified by
\citet{einasto_supercluster_2006}.  
According to our analysis, since the total luminosity of this FVS is
well beyond the threshold luminosity calibrated using the simulation,
these structures are candidates to merge and form a single virialized
system in the future.

\textbf{Bootes Superclusters}. 
The Bootes supercluster (SCL138 in \citet{einasto_optical_2001}) is
located at $z\simeq 0.065$, and Bootes-A (SCL150) lies directly behind
at $z\simeq 0.11$.
These superclusters have not been widely studied.  We find that a
single FVS of filamentary shape can be associated to each one of these
superclusters.

The comparison presented in this section helps visualize the
differences that can be found between different superstructure
catalogues; our approach has the relative advantage of allowing an
interpretation of known structures in terms of whether they will
evolve in the future into single virialized structures.
However, we remind the reader that this interpretation depends on the
assumed cosmology, which in the present case is the concordance
$\Lambda CDM$ model as well as the several hypothesis adopted.

\section{Clusters and Groups within FVS}  \label{S_SCSinFBS}

Although it has been claimed that the clustering properties of dark
matter haloes depend exclusively on their total mass, recent analyses
of numerical simulations have shown that they can be strongly affected
by their assembly history \citep[see e.g.][ and references
therein]{lacerna_nature_2010}.
If galaxy groups and clusters within superstructures have undergone a
different evolution, it would be expected systematic differences are
to be expected at in the present time.  The following analysis could
help to address the role of the large-scale structure in the formation
and evolution of galaxies.
In this section we derive a simple statistical analysis concerning
groups of galaxies and the FVS.
To this end, we use three different samples of galaxy systems:
(i) Groups of galaxies identified in the SDSS--DR7 (see
\sref{S_data}), suitable to search for correlations between halo mass
and the large scale environment they inhabit,
(ii) Abell ACO cluster catalog
\citep{andernach_cluster_1991}, that provides redshift measurements of
1059 clusters in the footprint area of the SDSS--DR7, and
(iii) X--ray clusters of galaxies, from the RASS catalogue
\citep{popesso_rass_2004} providing a suitable sample of bright X--ray
clusters within the SDSS area.

\begin{figure}
   \centering
   \includegraphics[width=0.5\textwidth]{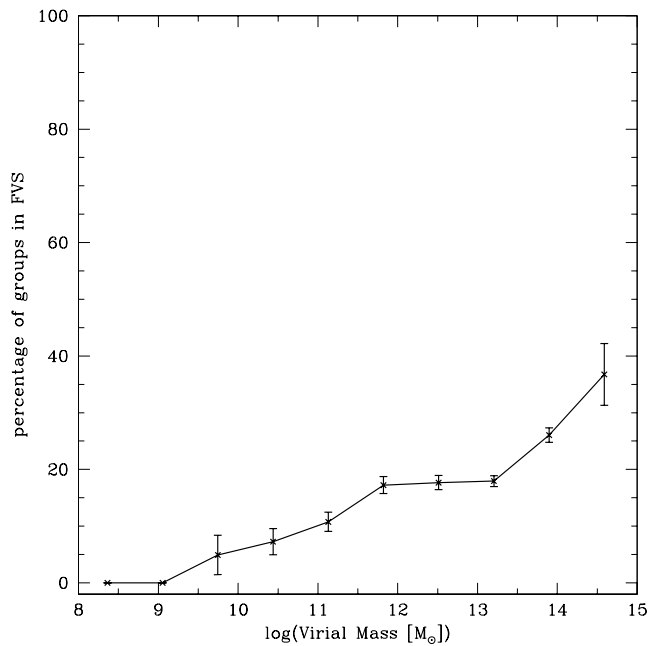}
   \caption{Percentage of DR7 galaxy groups inside FVS with
   respect to the total number of groups into virial mass bins.
   The first and second mass bins comprise 21 galaxy groups,
   although none of them 
   are located into a Future Virialized Structure.
   }
   \label{fig:GroupsFBS}
\end{figure}

We have computed the fraction of galaxy groups/clusters within FVS as
a function of their mass (\fref{fig:GroupsFBS}).
In this figure, error bars indicate Poisson uncertainties given by,

\begin{equation}
\sigma(M_{vir}^{(i)}) = \sqrt{N_{inFVS}^{(i)}}/N_{all}^{(i)},
\end{equation}

\noindent where $N_{inFVS}^{(i)}$ is the number of groups/clusters
inside a FVS in the \mbox{$i$-th} mass bin, and $N_{all}^{(i)}$ is the
total number of groups in the same mass bin.
As can be appreciated, the fraction of groups within FVS is greater
for higher group masses.
This result is also present in the catalogue of Abell--ACO clusters.
This catalogue supplies 177 clusters within the coverage area of
SDSS--DR7, out of which 119 lie within the supercluster sample defined
by \citet{einasto_optical_2001} and 63 lie within FVS.
A total number of 55 Abell clusters are found in both catalogues.
We compute fraction of Abell clusters within FVS as a function of the
richness class parameter.
We find only 14.5 per cent of type 0 clusters within FVS;
a value that increases for richer Abell clusters:
24.4 per cent for type 1, 34.4 per cent for type 2 and 41.7 per cent
for type 3 clusters.
Thus, there is a significantly larger probability of the richest Abell
clusters to be in FVS, in contrast to poor clusters.

\begin{figure}
   \centering
   \includegraphics[width=0.5\textwidth]{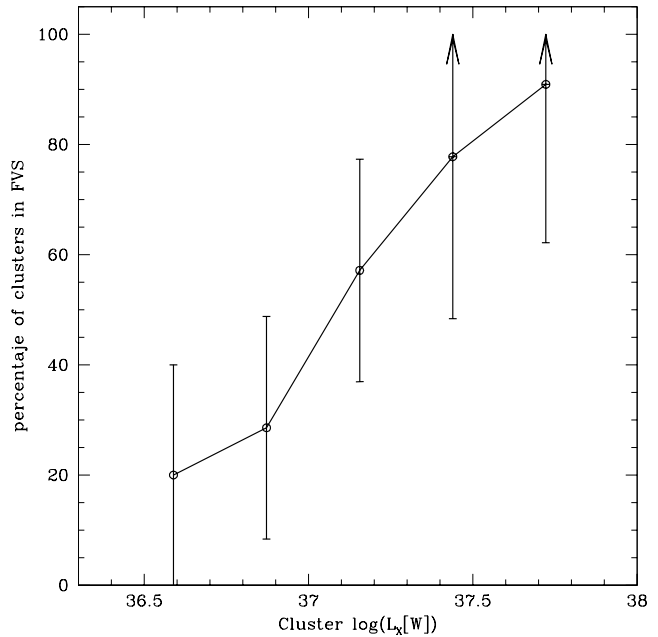}
   \caption{Percentage of X--ray clusters in FVS with
   respect to the total number of X--ray clusters, as a function
   of X--ray luminosity.
   }
   \label{fig:XrayFBS}
\end{figure}

It is also worth studying X--ray emitting clusters of galaxies.
While Abell clusters have an estimate of their richness that is
discrete and approximate, the X--ray cluster catalogue provides X--ray
luminosities, known to be good indicators of the underlying
gravitational potential well \citep{rykoff_LXM_2008}.
The catalogue of X--ray clusters also allows to study the fraction 
of clusters inside FVS as a function of the X--ray luminosity.
We obtain the same trend in this fraction as in the case of DR7
groups, as it can be seen in \fref{fig:XrayFBS}.
However, while the fraction of systems inside FVS is at most 0.3 for
the DR7 groups (except for the last virial mass bin, which has a large
uncertainty), we find that 85 per cent of the clusters with
\mbox{$L_X>4\,10^{37}$ W} are part of a FVS.
A similar value is found for superclusters by
\citet{einasto_optical_2001} using a compilation of X--ray clusters
based on the ROSAT All Sky Survey. 
The authors also found that in rich and very rich superclusters the
fraction of X--ray clusters is higher than the fraction of Abell
clusters, consistent with our findings.

The results of this section are in agreement with the expectation of
an assembly bias scenario, where the oldest, most massive systems are
preferentially located in present-day overdense regions. Also, this
suggests that the most frequent future cluster merger events will be
associated to the most massive clusters within FVS.


\section{Discussion}
\change
On the observational side, catalogues of superstructures have been presented on increasingly 
larger galaxy datasets (see \citeauthor{liivamagi_sdssdr7_2010}, \citeyear{liivamagi_sdssdr7_2010}, and references therein).
These catalogues provide important characterizations of the largest scale
structures and allow for different tests of structure formation
models. 
Regarding the evolution of these structures from the present, numerical
simulations reveal the conditions by which these systems would evolve
into isolation and dynamical equilibrium
within the $\Lambda$CDM scenario.
\citet{Busha_2005} consider the long term evolution of large structures, up
to $a=100$, using numerical simulations, and analyse the definition of
a radius that encloses all the mass that ultimately will form an
isolated structure.
The authors find that a transition region is found during the matter
dominated era ($a<a_{eq}$), between an inner hydrostatic region and an
outer region that expands with the perturbed Hubble flow. 
In this intermediate region, accretion of matter towards the mass
concentration is complex and gives rise to a variety of definitions of
scales to describe clusters.
At later times, the accretion region narrows and the
hydrostatic and turnaround scales merge to form a single zero-velocity
surface that unambiguously defines the halo mass.
\citet{araya-melo_future_evolution} argue that, once $\Lambda$ has
started to dominate the expansion of the universe, the cosmic web
growth stops, so that the spatial distribution of superclusters is
essentially the same starting from the present epoch.
The authors identify superclusters on the basis of criteria presented
in \citet{duenner_limits_2006}.  
These criteria, based on the spherical collapse model and implemented
on N-body simulations, allow to develop a method to isolate
structures in the present universe in redshift-space that will eventually
form a bound structure.
%
In a subsequent work, \citet{dunner_redshift_space_limits} presents a
detailed method to be applied in observational data. This method is
based on the assumption of spherical collapse and makes use of
projected velocity envelope templates and mass profiles.

The ''gravitationally bound" criterion is widely used, and was also
applied to our nearby Universe. 
\citet{Nearby_universe_Nagamine_2002} performed a numerical simulation
with initial conditions at $z=0$ reproducing the observed galaxy
distribution in the local Universe.
They conclude that the mass overdensity of the Local Group is above
the required threshold to be a bound structure in the future.
According to the authors, Andromeda and the Milky Way will probably
merge, while our Local Group and the Virgo Cluster will not form a 
virialized structure in the future.

From a theoretical point of view, \citet{einasto_2010_wavelet}
decompose the luminosity density field of
\citet{liivamagi_sdssdr7_2010} using a wavelet analysis. They also
study the formation of the cosmic web based on the evolution of the
density perturbation phases in numerical simulations.
They state that very rich superclusters are the result of large scale
($\lambda \geq 32h^{-1}Mpc$) density waves, combined in similar
phases. Also, the cosmic web develops in early stages of the Universe
evolution, and it is generated by the phase synchronization.

\begin{figure}
   \centering
   \includegraphics[width=0.5\textwidth]{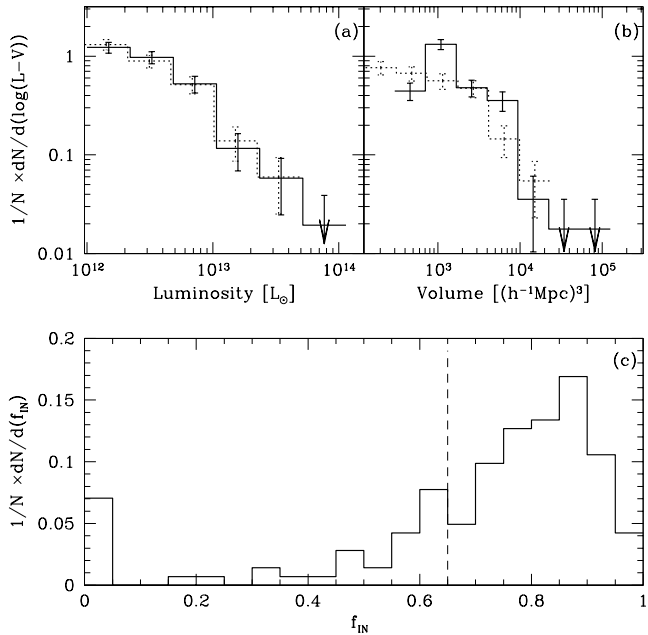}

   \caption{Upper panel: luminosity (a ) and volume (b) for FVS obtained from
the S2 sample (solid line) and for Superclusters in the catalogue of
\citet{liivamagi_sdssdr7_2010} (dotted line).  Error bars are Poisson
uncertainties within each bin.  Lower panel: Fraction of galaxies in
Superclusters that also belong to a FVS, $f_{IN}$.  
80 per cent of superclusters have at least a 65\% of their galaxies within FVS,
represented by the vertical dashed line. }

   \label{fig:liiv_vs_fvs}
\end{figure}

The largest supercluster catalogue constructed so far with SDSS-DR7
data \citep{liivamagi_sdssdr7_2010} allows to test our procedure and
compare the results.
As they apply the density field method using a wide range of threshold
density parameters (analogues to our $D_c$), we can directly contrast
the final results of both catalogues for the same value of $D_c=5.5$.
The \citet{liivamagi_sdssdr7_2010} catalogue is deeper, so we restrict
the analysis to $z=0.12$, the limit of our main sample of FVS.
Up to this distance, we identify 150 FVS compared to 142 superclusters
in \citet{liivamagi_sdssdr7_2010} catalogue.  In the upper panel of
\fref{fig:liiv_vs_fvs} we show luminosity and volume of these
structures. 
From the available list of galaxies in superclusters of
\citet{liivamagi_sdssdr7_2010}, we selected a subsample of 83.798
galaxies, with the same limits of our sample S2 ($0.04<z<0.12$ and
$M_r<-20.47$).
We find that 79.4\% of the galaxies in superclusters also belong to a
FVS, and only the 3.4\% of galaxies outside of superclusters are
associated to a FVS.
We also analyse the galaxies within each supercluster separately,
defining the fraction of galaxies in the supercluster that are part of
a FVS $f_{IN}$, and the fraction of galaxies in the supercluster that
do not belong to any FVS, $f_{OUT}$.

In the lower panel of \fref{fig:liiv_vs_fvs} we show the $f_{IN}$
fraction for the galaxies within each supercluster. We find that 80\%
of superclusters have $f_{IN}$ over 0.65, and $f_{OUT}$ lower than
0.35.
The previous analyses indicate that the general features of the
structures of \citet{liivamagi_sdssdr7_2010} and those of the FVS show
a very good agreement.

\changeend

\section{Conclusions} \label{s_conclusus} 

The properties of the largest structures in the Universe provide
important information on the cosmological model.
In this work we have developed a physically motivated procedure to
isolate structures in the nearby universe that will evolve into
gravitationally virialized systems in the future, within a $\Lambda
CDM$ framework.
The method starts from a smoothed luminosity density map, where high
density peaks are separated as superstructure candidates.
This method depends on a number of free parameters that we determined
on a theoretical basis.
We adopted an Epanechnikov Kernel following recent similar
implementations of the method
\citep[e.g.][]{einasto_superclusters_2007,costaduarte_morphological_2010}.
We then calibrated the luminosity overdensity parameter $D_T$
according to results from \citet{duenner_limits_2006}, who used
numerical simulations to establish the conditions for a superstructure
to be virialized at $a\sim100$.
We find that a cut in the total luminosity of superstructures at
$L=10^{12} L_{\odot}$ is convenient to additionally clean the sample of
FVS (see \fref{fig:dlcs}).
This is due to the fact that although some superstructures have a
total luminosity overdensity above the critical value, their mass
can still be not enough to ensure the future collapse onto a
virialized system.
We used a mock catalogue to further constrain and test the reliability
of the identification method.
We defined three samples of FVS
from volume limited samples of galaxies.
From the comparison of these samples we conclude that the method is
not affected by the constraints in the survey redshift and the
luminosity of galaxies, since the small differences can be explained
by the corrections implemented assuming a universal luminosity
function.

The joint analysis of group catalogues and the samples of FVS indicate
that the fraction of groups within FVS increases with the virial mass
(see \fref{fig:GroupsFBS}).
The fraction of X--ray clusters belonging to FVS is significantly
larger, and also increases with the X--ray luminosity
(\fref{fig:XrayFBS}).
This is consistent with the assembly bias scenario, with massive, old assembled systems 
located preferentially in present-day superstructures as FVS.
Thus, future cluster merger events 
will often involve the most massive, bright X--ray clusters.
We have also analysed particular superclusters previously studied by
other authors.
Many known superclusters are found to be coincident with one or more
FVS in our sample.
As an example, our analysis shows that the
Ursa Major supercluster is composed by four filaments,
each one a distinct FVS; and the
Corona Borealis and Bootes superclusters are clearly associated with
FVS in our catalogue.



\section*{acknowledgements}
We thank the referee, Jaan Einasto, for his through review and highly 
appreciate the comments and suggestions, which greatly improved this work.
This work was partially supported by the
Consejo Nacional de Investigaciones Cient\'{\i}ficas y T\'ecnicas
(CONICET), and the Secretar\'{\i}a de Ciencia y Tecnolog\'{\i}a,
Universidad Nacional de C\'ordoba, Argentina.
Funding for the SDSS and SDSS-II has been provided by the Alfred P.
Sloan Foundation, the Participating Institutions, the National Science
Foundation, the U.S. Department of Energy, the National Aeronautics
and Space Administration, the Japanese Monbukagakusho, the Max Planck
Society, and the Higher Education Funding Council for England. The
SDSS Web Site is http://www.sdss.org/.
The SDSS is managed by the Astrophysical Research Consortium for the
Participating Institutions. The of the Royal Astronomical Society
Participating Institutions are the American Museum of Natural History,
Astrophysical Institute Potsdam, University of Basel, University of
Cambridge, Case Western Reserve University, University of Chicago,
Drexel University, Fermilab, the Institute for Advanced Study, the
Japan Participation Group, Johns Hopkins University, the Joint
Institute for Nuclear Astrophysics, the Kavli Institute for Particle
Astrophysics and Cosmology, the Korean Scientist Group, the Chinese
Academy of Sciences (LAMOST), Los Alamos National Laboratory, the
Max-Planck-Institute for Astronomy (MPIA), the Max-Planck-Institute
for Astrophysics (MPA), New Mexico State University, Ohio State
University, University of Pittsburgh, University of Portsmouth,
Princeton University, the United States Naval Observatory, and the
University of Washington.   
The Millennium Run simulation used in this paper was carried out by the
Virgo Supercomputing Consortium at the Computer Centre of the
Max--Planck Society in Garching.  The semi--analytic galaxy catalogue
is publicly available at
http://www.mpa-garching.mpg.de/galform/agnpaper.

\bibliographystyle{mn2e}
\bibliography{Bibliography}

\end{document}